\documentclass[manuscript]{acmart}

\AtBeginDocument{%
  \providecommand\BibTeX{{%
    \normalfont B\kern-0.5em{\scshape i\kern-0.25em b}\kern-0.8em\TeX}}}

\setcopyright{acmcopyright}
\copyrightyear{2018}
\acmYear{2018}
\acmDOI{10.1145/1122445.1122456}

\acmConference[Woodstock '18]{Woodstock '18: ACM Symposium on Neural
  Gaze Detection}{June 03--05, 2018}{Woodstock, NY}
\acmBooktitle{Woodstock '18: ACM Symposium on Neural Gaze Detection,
  June 03--05, 2018, Woodstock, NY}
\acmPrice{15.00}
\acmISBN{978-1-4503-XXXX-X/18/06}

\usepackage{subcaption}
\usepackage{graphicx}
\usepackage{caption}
\begin{document}

\title{AutoRec: An Automated Recommender System}
\author{Ting-Hsiang Wang}
\email{thwang1231@tamu.edu}
\affiliation{%
  \institution{Texas A\&M University}
  \state{Texas}
  \country{United States}
}

\author{Qingquan Song}
\email{song_3134@tamu.edu}
\affiliation{%
  \institution{Texas A\&M University}
  \state{Texas}
  \country{United States}
}

\author{Xiaotian Han}
\email{han@tamu.edu}
\affiliation{%
  \institution{Texas A\&M University}
  \state{Texas}
  \country{United States}
}

\author{Zirui Liu}
\email{tradigrada@tamu.edu}
\affiliation{%
  \institution{Texas A\&M University}
  \state{Texas}
  \country{United States}
}

\author{Haifeng Jin}
\email{jin@tamu.edu}
\affiliation{%
  \institution{Texas A\&M University}
  \state{Texas}
  \country{United States}
}

\author{Xia Hu}
\email{hu@cse.tamu.edu}
\affiliation{%
  \institution{Texas A\&M University}
  \state{Texas}
  \country{United States}
}

\renewcommand{\shortauthors}{Wang, et al.}

\begin{abstract}
  Realistic recommender systems are often required to adapt to ever-changing
  data and tasks or to explore different models systematically.
  To address the need, we present \textbf{AutoRec}~\footnote{\hspace{0.1cm}
  AutoRec GitHub:\hspace{0.1cm} https://github.com/datamllab/AutoRecSys}~\footnote{\hspace{0.1cm}
  AutoRec Video:\hspace{0.1cm} https://www.youtube.com/watch?v=z0HkKGVAQkE}, an
  open-source automated machine learning (AutoML) platform extended from the
  TensorFlow ecosystem and, to our knowledge, the first framework to leverage
  AutoML for model search and hyperparameter tuning in deep recommendation models.
  AutoRec also supports a highly flexible pipeline that accommodates both sparse
  and dense inputs, rating prediction and click-through rate (CTR) prediction tasks,
  and an array of recommendation models.
  Lastly, AutoRec provides a simple, user-friendly API.
  Experiments conducted on the benchmark datasets reveal AutoRec is reliable and
  can identify models which resemble the best model without prior knowledge.
\end{abstract}

\maketitle

%A bit different is that, AutoRec can directly do search, but Gluon-CV and Gluon-nlp just implement task-specific models, and AutoGluon can borrow them for search purpose.

\section{Introduction}
  Most recommender systems are highly specialized to handle specific data
  and tasks.
  For example, NCF~\cite{ncf} takes user-item implicit feedback data as
  inputs for the rating prediction task;
  and DeepFM~\cite{deepfm} leverages both numerical and categorical data
  for the CTR prediction task.
  However, high degree of specialization comes at the expense of model
  adaptability and tuning complexity.
  As recommendation tasks evolve over time and additional types of data are
  collected, the originally apt model can either become obsolete or require
  tremendous tuning efforts.
  So far, several pipelines for recommender systems, e.g.,
  OpenRec~\cite{openrec} and SMORe~\cite{smore}, tried to address the
  adaptability issue via providing modular base blocks that can be selected
  according to the context of recommendation.
  Nevertheless, both determining the blocks to use and tuning the model
  parameters are not straightforward when facing new data and changing
  tasks.
  
  In order to bridge the gap, we present AutoRec, which aims to provide
  an end-to-end solution to automate model selection and hyperparameter
  tuning.
  While many AutoML libraries, such as Auto-Sklearn~\cite{autosklearn} and
  TPOT~\cite{tpot} have shown promising results in general-purpose machine
  learning tasks (e.g., regression and hyperparameter tuning) and our
  fruitful efforts with AutoKeras~\cite{autokeras} extended AutoML
  to multi-modal data and multi-task training (e.g., text and image
  classificaiton), few models incorporate AutoML for recommendation tasks.
  And for the few which do, their approaches are often too narrow for general
  recommendation models.
  For example, AutoInt~\cite{autoint} and AutoCTR~\cite{autoctr} focus on
  searching interactions for only CTR prediction task.
  Hence, one major novelty of AutoRec is its modular, searchable pipeline
  architecture tailored for recommendation models.
  
\begin{figure}
  \centering
  \begin{minipage}{.45\columnwidth}
    \includegraphics[width=\textwidth]{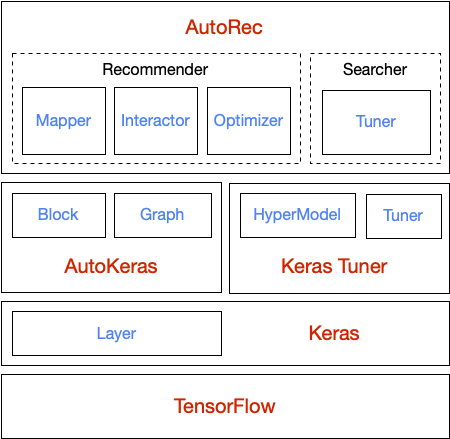}
    \caption{\normalsize AutoRec architecture and the TensorFlow ecosystem. Notice red (blue) indicates package (file).}
    \label{fig:architecture}
  \end{minipage}
  \hfill
  \begin{minipage}{.5\columnwidth}
    \includegraphics[width=\textwidth]{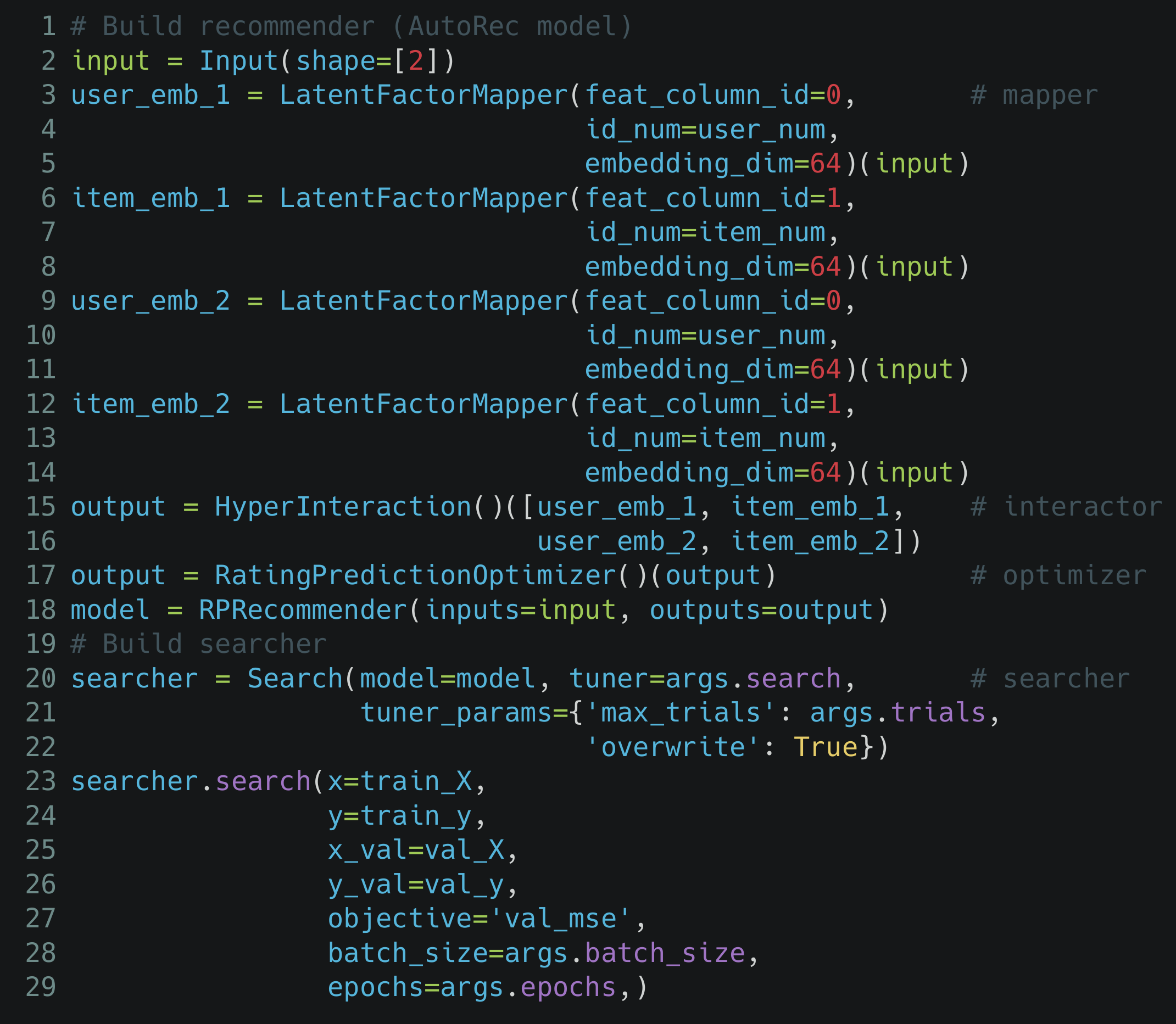}
    \caption{\normalsize Building a searchable recommendation model for the rating prediction task (AutoRec-RP).}
    \label{fig:build}
  \end{minipage}
\end{figure}

\section{AutoRec Architecture Overview}

\subsection{Recommender Construction}
  As shown in Figure~\ref{fig:architecture}, a recommender is composed of
  3 kinds of specialized Blocks: \textit{mapper}, \textit{interactor},
  and \textit{optimizer}, and the Graph is responsible for putting them
  together to form a recommendation model's search space.
  The HyperModel is the basis for Block.
  And the Tuner is the basis for Searcher's tuning algorithms.
 
 \textbf{Mapper.} This base block converts data features into
  low-dimensional latent factors (embeddings) so different entities can be
  compared numerically.
  The numerical/dense features (e.g., time, rating, listening count) are
  naturally comparable and thus directly mapped as embeddings.
  For categorical/sparse features (e.g., devise, item category, and
  user and item identifiers), the preprocessor must fit and transform
  the features before the mapper can be applied.
  Figure~\ref{fig:build} line 3 to 14 shows the declaration of
  multiple mappers.

  \textbf{Interactor.} This base block simulates different ways of
  interactions between entities.
  Currently, the 7 pre-structured interactions are as follows:
  \textit{MLPInteraction}, \textit{ConcatenateInteraction},
  \textit{FMInteraction} , \textit{CrossNetInteraction},
  \textit{SelfAttentionInteraction}, \textit{ElementwiseInteraction}, and
  \textit{RandomSelectInteraction}, and each of them have different
  tunable hyperparameters.
  In contrast, the eighth interactor, \textit{HyperInteraction}, is
  unstructured and supports both hyperparameter tuning and model search.
  Figure~\ref{fig:build} line 15 to 16 shows how multiple mappers
  are chained into an interactor.
 
  \textbf{Optimizer.} This base block specifies how to generate predicted
  values and how to measure the deviation between the predicted value and
  the ground truth, i.e., rating for the rating prediction task and label
  for the CTR prediction task.
  Figure~\ref{fig:build} line 17 shows how the interactor is chained
  into an optimizer.

\subsection{Searcher Construction} 
  The automated search stage generates a searcher object, which is
  composed of 3 types of tuner algorithms: \textit{RandomSearch},
  \textit{Greedy}, and \textit{BayesianOptimization}.
  To interact with the searcher, simply pass the recommender object
  obtained from the recommender construction stage, as shown in
  Figure~\ref{fig:build} line 20 to line 29.
  
\section{Evaluation}
\subsection{Framework Simplicity and Adaptability}
  The example function calls for the AutoRec is shown in Figure~\ref{fig:build}.
  As we can see, the entire process can be done by chaining
  \textit{declaration calls} and \textit{execution calls}.
  Using this simple interface, we show the searchable version of many
  mainstream recommendation models, such as DeepFM~\cite{deepfm},
  DLRM~\cite{dlrm}, AutoInt~\cite{autoint}, CrossNet~\cite{crossnet},
  MF~\cite{mf}, NCF~\cite{ncf}, and MLP, can all be assembled with a few
  lines of code.

\begin{table}[]
  \caption{AutoRec platform performance for the click-through rate prediction task (logloss).}
  \small
  \scalebox{0.7}{
  \begin{tabular}{|c|c|c|c|c|c|c|c|c|c|c|c|c|c|c|c|}
    \hline
        Model       & \multicolumn{3}{c|}{DeepFM} & \multicolumn{3}{c|}{DLRM}   & \multicolumn{3}{c|}{CrossNet}        & \multicolumn{3}{c|}{AutoInt} & \multicolumn{3}{c|}{AutoRec-CTR}      \\
    \hline
        Tuner       & Random & Greedy & Bayesian & Random & Greedy & Bayesian & Random          & Greedy & Bayesian & Random  & Greedy & Bayesian & Random          & Greedy & Bayesian \\
    \hline
        Avazu 500K  & 0.4064 & 0.4012 & 0.4025   & 0.4014 & 0.4020 & 0.4042   & \textbf{0.3992} & 0.3995 & 0.4028   & 0.4033  & 0.4054 & 0.4089   & \underline{0.3995}   & 0.4047 & 0.4093   \\
        Criteo 500K & 0.4811 & 0.4795 & 0.4722   & 0.4759 & 0.4743 & 0.4783   & \textbf{0.4713} & 0.4786 & 0.4761   & 0.4755  & 0.4718 & 0.4739   & \underline{0.4747}   & 0.4762 & 0.4714   \\
    \hline
  \end{tabular}
  }
  \label{tab:ctr}
\end{table}

  \begin{table}[]
  \caption{AutoRec platform performance for the rating prediction task (MSE loss).}
  \small
  \scalebox{0.7}{
  \begin{tabular}{|c|c|c|c|c|c|c|c|c|c|c|c|c|}
    \hline
        Model           & \multicolumn{3}{c|}{MF}    & \multicolumn{3}{c|}{MLP}   & \multicolumn{3}{c|}{NCF}   & \multicolumn{3}{c|}{AutoRec-RP} \\
    \hline
        Tuner           & Random & Greedy & Bayesian & Random & Greedy & Bayesian & Random & Greedy & Bayesian & Random             & Greedy  & Bayesian \\
    \hline
        Movielens 1M    & 0.7550 & 0.7502 & 0.7517   & 0.7681 & 0.7706 & 0.7597   & 0.7720 & 0.7520 & 0.7723   & \underline{0.7497} & 0.7510  & \textbf{0.7494}   \\
        Netflix         & 0.7478 & 0.7402 & 0.7287   & 0.7553 & 0.7652 & 0.7552   & 0.7063 & 0.6440 & 0.7065   & \textbf{0.6371}    & 0.7401  & \underline{0.6453}   \\
    \hline 
  \end{tabular}
  }
  \label{tab:rp}
\end{table}

\subsection{Recommendation Performance}
  We use 500K data from 
  Avazu dataset~\cite{avazu} and Criteo dataset~\cite{criteo} for the CTR prediction
  task, and we use the complete Movielens 1M dataset~\cite{movielens} and
  Netflix dataset~\cite{netflix} for the rating prediction task.
  The train-validation-test ratio is 8:1:1.
  The training parameters are: epoch=10, dimension=64, early stop=1, and trial=10.
  Due to its large size, Netflix dataset has batch size set to 512000, while others
  have batch size set to 1024.
  
  Table~\ref{tab:ctr} and Table~\ref{tab:rp} show the performance of AutoRec in terms
  of hyperparameter tuning and model search.
  Specifically, AutoRec tunes the hyperparameter for all models, with AutoRec-CTR and
  AutoRec-RP further subject to model search.
  We remark all models found by AutoRec in both tasks yielded satisfialbe results by
  experience, verifying the platform's reliability.
  In Table~\ref{tab:ctr}, AutoRec-CTR is able to find the model whose performance is the
  closest to the best model, CrossNet, testifying AutoRec's modeling ability without
  prior knowledge.
  In Table~\ref{tab:rp}, AutoRec-RP simply identifies the best models with different
  tuner algorithms, and its other findings are also very promising.

\section{Conclusion}
  Realistic recommender systems often need to adapt to ever-changing scenarios or to
  explore options systematically, and AutoML fits right into the spot.
  In this demonstration, we present AutoRec, an open-source AutoML platform for
  recommendation tasks based on the TensorFlow ecosystem.
  AutoRec supports both hyperparameter tuning and model search, and our experiments
  verify it can identify close-to-the-best model without prior knowledge.
  In the near future, we plan to update AutoRec for larger search space and
  comprehensive plug-and-play examples.
  In addition, we look forward to your participation in this open-source project.

\bibliographystyle{ACM-Reference-Format}
\bibliography{paper} 

\end{document}